\documentclass{PoS}

\usepackage{amsmath,amssymb}
\definecolor{red}{rgb}{1,0,0}

\graphicspath{{./figures/}}

\def\eq#1{(\ref{#1})}

\def\Eq#1{Eq.~(\ref{#1})}
\def\Fig#1{Fig.~\ref{#1}}
\def\Sec#1{Sec.~\ref{#1}}

\def\tr{\text{tr}}

\title{Strangeness Neutrality and the QCD Phase Diagram}

\ShortTitle{Strangeness Neutrality and the QCD Phase Diagram}

\author{\speaker{Fabian Rennecke}\\
        Physics Department, Brookhaven National Laboratory, Upton, NY 11973, USA\\
        E-mail: \email{frennecke@bnl.gov}}

\author{Wei-jie Fu\\
       School of Physics, Dalian University of Technology, Dalian, 116024, P.R.\ China\\
        E-mail: \email{wjfu@dlut.edu.cn}}
        
\author{Jan M.\ Pawlowski\\
       Institut für Theoretische Physik, Universität Heidelberg, Philosophenweg 16, 69120 Heidelberg, Germany\\
        E-mail: \email{j.pawlowski@thphys.uni-heidelberg.de}}

\abstract{
Since the incident nuclei in heavy-ion collisions do not carry strangeness, the global net strangeness of the detected hadrons has to vanish. We show that there is an intimate relation between strangeness neutrality and baryon-strangeness correlations. In the context of heavy-ion collisions, the former is a consequence of quark number conservation of the strong interactions while the latter are sensitive probes of the character of QCD matter. We investigate the sensitivity of baryon-strangeness correlations on the freeze-out conditions of heavy-ion collisions by studying their dependence on temperature, baryon- and strangeness chemical potential. The impact of strangeness neutrality on the QCD equation of state at finite chemical potentials will also be discussed. We model the low-energy sector of QCD by an effective Polyakov loop enhanced quark-meson model with 2+1 dynamical quark flavors and use the functional renormalization group to account for the non-perturbative quantum fluctuations of hadrons.
}

\FullConference{Corfu Summer Institute 2018 "School and Workshops on Elementary Particle Physics and Gravity"\\
		(CORFU2018)\\
		31 August - 28 September, 2018\\
		Corfu, Greece}

\begin{document}

\section{Introduction}

Understanding the phase structure of the strong interactions, described by Quantum Chromodynamics (QCD), is a key cornerstone for our understanding of the formation of matter in the universe. Phase transitions from states of matter described by the fundamental degrees of freedom of QCD, quarks and gluons, to states characterized by hadronic matter are believed to have occurred in the early universe only microseconds after the big bang, and potentially also in the interior of neutron stars. Recreating the extreme conditions necessary for a systematic study of the phase diagram requires the collisions of heavy ions at ultrarelativistic energies, see e.g.\ \cite{Gazdzicki:995681, BESwp, Friman:2011zz, Kekelidze:2016hhw, GALATYUK201441, Sako:2014fha}. By varying the beam-energy $\sqrt{s}$, it is possible to scan different regions of the phase diagram in these heavy-ion collisions.

One mayor challenge is to experimentally identify phase transitions in the first place. At low baryochemical potential $\mu_B$, it is known that there is a continuous crossover rather than a proper phase transition \cite{Aoki:2006we, Bellwied:2015rza, Bazavov:2018mes}. Effective model calculations suggest that this crossover might turn into a critical endpoint (CEP), where the transition is of second order, followed by a first-order phase transition \cite{Stephanov:2007fk}. However, since only hadrons can be measured directly, the challenge is to identify observables which inherit signatures of a crossover or a phase transition that might have occurred during the evolution of the quark-gluon plasma created in heavy-ion collisions. The CEP is of particular interest here, since the correlation length diverges at a second-order phase transition, giving rise to critical phenomena. A prominent example for such observables are non-Gaussian event-by-event fluctuations of particle multiplicities \cite{Stephanov:1999zu, Stephanov:2008qz}.

The situation is additionally complicated by the fact that the QCD phase diagram is not only spanned in the two-dimensional plane of temperature $T$ and $\mu_B$, but is a multi-dimensional object depending on additional directions, such as strangeness chemical potential $\mu_S$, charge chemical potential $\mu_Q$, magnetic fields and rotation frequency. Additionally, non-equilibrium effects related to the expansion- and cooling rate of the plasma as well as strong electrical fields could modify the signatures of the phase structure. In general, the hadrons reaching the detector inherit the properties of the QCD medium at freeze-out. There is compelling evidence that at least some of the properties of the system at freeze-out can be described by equilibrium thermodynamics \cite{BraunMunzinger:2003zd}. Furthermore, the typical freeze-out time is many orders of magnitude shorter than the timescale of flavor-changing weak decays. Hence, quark number conservation of the strong interactions has to be taken into account. Each quark flavor therefore has a chemical potential associated to it. Since the typical energy scales for the study of the phase diagram are $2 \pi T_c \approx 1 \,\text{GeV}$, where $T_c$ is the (pseudo-) critical temperature, we can focus on the three lightest flavors $q = (u, d ,s)^T$, and the corresponding chemical potentials can be described by the matrix,
\begin{align}\label{eq:mumat}
\mu =
 \begin{pmatrix}
\mu_u & 0 & 0 \\ 0 & \mu_d & 0 \\ 0 & 0 & \mu_s
\end{pmatrix}
=
 \begin{pmatrix}
\frac{1}{3} \mu_B + \frac{2}{3} \mu_Q & 0&0 \\
0& \frac{1}{3} \mu_B - \frac{1}{3} \mu_Q &0 \\
0&0& \frac{1}{3} \mu_B - \frac{1}{3} \mu_Q - \mu_S
\end{pmatrix}\,.
\end{align}
The coupling to the quarks is given by $\bar q \gamma_0 \mu q$.
With these chemical potentials, the particle number correlations mentioned above can be described by the generalized susceptibilities,
\begin{align}\label{eq:sus}
  \chi_{ijk}^{BSQ}(T,\mu_B,\mu_S,\mu_Q) = \frac{\partial^{i+j+k}
  p(T,\mu_B,\mu_S,\mu_Q)/T^4}{\partial\hat\mu_B^i\partial\hat\mu_S^j\partial\hat\mu_Q^k}\,,
\end{align}
where $p$ is the pressure of QCD and $\hat\mu = \mu/T$. Owing to quark number conservation, the associated particle numbers, net-baryon number, net-strangeness and net-charge,
\begin{align}\label{eq:nets}
\langle B \rangle = \langle N_B - N_{\bar B} \rangle = V T^3 \chi_{1}^{B}\,,
 \quad \langle S \rangle = \langle N_{\bar S} - N_S \rangle = V T^3 \chi_{1}^{S}\,,
  \quad \langle Q \rangle = \langle N_Q - N_{\bar Q} \rangle = V T^3 \chi_{1}^{Q}\,,
\end{align}
are fixed by the initial conditions of the heavy-ion collision. Since the incident nuclei have vanishing net strangeness, the \emph{strangeness neutrality} condition $\langle S \rangle = 0$ therefore has to be fulfilled during each stage of the collision. The net-charge is fixed by the charge of the nuclei. Since the net-baryon number at mid-rapidity depends on the beam-energy, we assume that $\mu_B$ is a free parameter here, which can be mapped one-to-one to a given beam-energy, see e.g.\ \cite{Adamczyk:2017iwn}.

In this talk, we address the question how the conservation of net-strangeness influences the phase structure and thermodynamics of QCD. For simplicity, we restrict ourselves to $\mu_Q = 0$ for the most part. Note that realistic charge chemical potentials are indeed significantly smaller than the strangeness chemical potential \cite{Bazavov:2012vg}. We first derive a simple relation between strangeness neutrality and baryon-strangeness correlations $C_{BS}$. This elucidates the interplay between the composition of QCD matter at different $T$ and $\mu$ and particle number conservation. We demonstrate that $C_{BS}$ carries signatures of the chiral phase transition at finite $\mu_B$. This is based on \cite{Fu:2018swz}. Then, we show how thermodynamic properties and the chiral and deconfinement phase transitions are affected by strangeness neutrality. These results are based on \cite{Fu:2018qsk}.

\section{Strangeness neutrality}\label{sec:strn}

As already mentioned, particle number conservation in heavy-ion collisions implies that the the net-strangeness has to vanish and net-charged is fixed from the incident nuclei. For simplicity, we focus on isospin-symmetric matter, $\mu_Q = 0$. This implies that up- and down-quarks can be treated equally and we will refer to them as light quarks $l = u = d$. Strangeness neutrality implicitly defines the strangeness chemical potential,
\begin{align}\label{eq:mus0}
\begin{split}
\mu_{S0}(T,\mu_B) =  \mu_S(T,\mu_B)\big|_{\langle S \rangle = 0}\,.
\end{split}
\end{align}
For a given $T$ and $\mu_B$ (or $\sqrt{s}$) it is always possible to find a strangeness chemical potential such that the net-strangeness vanishes. This is most easily seen in the QGP phase at very large $T \gg T_c$, where all strangeness is carried by the strange quarks and antiquarks. Since $\mu_B$ couples equally to all quark flavors, see \Eq{eq:mumat}, increasing $\mu_B$ will also increase the number of strange quarks in the system. Since, for historical reasons, a positive strangeness chemical potential increases antistrange over strange quarks (note the sign of $\mu_S$ in \Eq{eq:mumat}), the effect of $\mu_B$ can always be compensated by
\begin{align}\label{eq:mus0lim}
\mu_{S0}(T\gg T_c, \mu_B) = \frac{1}{3} \mu_B\,.
\end{align}
In the hadronic phase, this situation is complicated by the fact that both baryons and mesons can carry strangeness. This gives rise to a nontrivial $\mu_{S0}$ which sensitively depends on the composition of strongly interacting matter. Since the net-strangeness density, $n_S$, is given by $n_S = T^3 \chi_1^S$, see \Eq{eq:nets}, the definition of $\mu_{S0}$ implies
\begin{align}
\chi_1^S(T,\mu_B,\mu_{S0}) = 0\,.
\end{align}
By taking a total derivative with respect to $\mu_B$, the right-hand side is still zero and we find
\begin{align}
\begin{split}
\frac{d}{d\hat\mu_B}\chi_1^S(T,\mu_B,\mu_{S0}) = \chi^{BS}_{11}(T,\mu_B,\mu_{S0}) + \chi^{S}_{2}(T,\mu_B,\mu_{S0}) \frac{\partial \hat \mu_{S0}}{\partial \hat \mu_B}  = 0\,.
\end{split}
\end{align}
This provides an implicit definition of the strangeness chemical potential in terms of the differential equation,
\begin{align}\label{eq:muscbs}
\frac{\partial\mu_{S0}}{\partial\mu_B} = \frac{1}{3}\, C_{BS}(T,\mu_B,\mu_{S0})\,,
\end{align}
where we defined the baryon-strangeness correlation:
\begin{align}\label{eq:cbsdef}
C_{BS} = -3 \frac{\chi^{BS}_{11}}{\chi^{S}_{2}}\,.
\end{align}
Hence, by integrating the measured $C_{BS}$ over the beam-energy for isospin-symmetric matter, one could extract the strangeness chemical potential. Conversely, the slope of $\mu_{S0}$ in $\mu_B$-direction determines $C_{BS}$. Baryon-strangeness correlations have been introduced in \cite{Koch:2005vg} as a diagnostic to elucidate the nature of strongly interacting matter. In terms of particle number correlations, one has at strangeness neutrality:
\begin{align}\label{eq:cbscorr}
C_{BS} = -3 \frac{\langle BS \rangle}{\langle S^2 \rangle}\,.
\end{align}
If, as above, we assume we are deep in the deconfined phase with $T \gg T_c$ where the strong coupling is very small, there is no correlation between different quark flavors. Hence, only strange quarks can contribute to both the denominator and the numerator of \Eq{eq:cbscorr}. Furthermore, there is a strict relation between strangeness $S$ and the baryon number $B_s$ of strange quarks, $B_s = -S/3$. Hence,
\begin{align}\label{eq:cbslim}
C_{BS}(T\gg T_c, \mu_B) = 1\,.
\end{align}
Thus, $C_{BS}$ has to become unity deep in the deconfined phase. The same result follows from Eqs.\ \eq{eq:mus0lim} and \eq{eq:muscbs}. In the hadronic phase, baryons can carry, aside from baryon number, also strangeness, while mesons can only carry strangeness. Strange baryons give the dominant contribution to the numerator of \Eq{eq:cbscorr}, while both strange mesons and baryons are important contributions for the denominator. For small $\mu_B$, mesons significantly outnumber baryons and $C_{BS}$ is expected to be smaller than unity. For large $\mu_B$, but still in the confined phase, baryons gain in relevance. Positive strangeness is primarily carried by kaons and and negative strangeness by the hyperons $\Lambda$ and $\Sigma$. So at large $\mu_B$, the strangeness carried by kaons is compensated by the hyperons and $C_{BS}$ can become larger than unity. We see that $C_{BS}$ is determined by an intriguing interplay between the different degrees of freedom of QCD in the different regimes of the phase diagram.

This is the reason why, despite its simplicity, \Eq{eq:muscbs} is an intriguing relation: It gives precise meaning to the intuitive connection between the conditions necessary to realize particle number conservation (described by $\mu_{S0}$) and the composition of QCD matter, which is sensitive to its phase structure (described by $C_{BS}$). In the following, we want to explore this relation in more detail.

But before we do that, we want to note that \Eq{eq:muscbs} easily generalizes to the case of isospin-asymmetric matter with $\mu_Q \neq 0$. In this case, in addition to $\chi_1^S = 0$, one has the condition $\chi_1^Q/\chi_1^B = Z/A$, where $Z$ is the atomic number and $A$ is the mass number of the incident nuclei. For instance, both for Au and Pb, one has $Z/A \approx 0.4$. This yields the following generalization of \Eq{eq:muscbs}:
\begin{align}
\begin{split}
  \frac{\partial \mu_{S0}}{\partial \mu_B} &= 
                                             \frac{1}{3} C_{BS}
                                             -\frac{\chi_{11}^{QS}}{
                                             \chi_{2}^{S}}\frac{\partial\mu_{Q0}}{\partial\mu_B}\,,\\
  \frac{\partial \mu_{Q0}}{\partial \mu_B} &=
                                             \frac{\chi_{11}^{BS}\big(\chi_{11}^{SQ}-\frac{Z}{A}
                                             \chi_{11}^{BS}\big)-\chi_{2}^{S}\big(\chi_{11}^{BQ}-\frac{Z}{A}
                                             \chi_{2}^{B}\big)}{\chi_{2}^{S}\big(\chi_{2}^{Q}-\frac{Z}{A}\chi_{11}^{BQ}\big)
                                             -\chi_{11}^{SQ}\big(\chi_{11}^{SQ}-\frac{Z}{A}
                                             \chi_{11}^{BS}\big)}\,.
\end{split}
\end{align}
These differential equations generalize the relations used by lattice QCD simulations for the determination of the freeze-out conditions of heavy-ion collisions, which are only valid at small $\mu_B/T$, to arbitrary $\mu_B$ and $T$ \cite{Bazavov:2012vg, Borsanyi:2013hza, Bazavov:2017tot}.

\section{Low-energy QCD and the functional renormalization group}

The impact of strangeness neutrality at
  finite $\mu_B$ and $T$ is studied within a low-energy
  effective theory of QCD as put forward in \cite{Fu:2018qsk}. In order
to capture the main features of strangeness, the quantum-, thermal- and
density fluctuations of open strange mesons, strange baryons and
quarks have to be taken into account. Since kaons are pseudo-Goldstone
bosons of spontaneous chiral symmetry breaking, they are the most
relevant strange degrees of freedom in the mesonic sector. Moreover,
chiral symmetry dictates that if kaons are included as effective
low-energy degrees of freedom, all other mesons in the lowest scalar
and pseudoscalar meson nonets have to be included as well. Confinement of quarks is effectively taken into account by coupling quarks to a uniform temporal gluon background field
$\bar A_0 = \bar A_0^{(3)}\, t^3 + \bar A_0^{(8)}\, t^8$, with the color group generators
$t^c\in SU(3)$.
In total, this gives rise to a 2+1 flavor Polyakov-loop enhanced quark-meson (PQM) model with the Euclidean
effective action ($\beta = 1/T$):
\begin{align}\label{eq:ea}  
  \Gamma_k = \int_0^{\beta} \!\!dx_0 \int \!\!d^3x\Big\{
             \bar q \big(\gamma_\nu D_\nu+\gamma_\nu C_\nu\big)q + h\,
             \bar q\,\Sigma_5 q+\text{tr}\big(\bar D_\nu\Sigma\!\cdot\!\bar D_\nu
             \Sigma^\dagger\big)+\widetilde U_k(\Sigma) + U_\text{glue}(L,\bar L)
             \Big\}\,.      
\end{align}
Quantum, thermal and density fluctuations of modes with Euclidean
momenta $k \leq |p| \lesssim 1$ GeV have been integrated out. We assume that the gauge degrees of freedom are fully integrated out at this scale. The
gauge covariant derivative only contains the background gauge field,
$D_\nu = \partial_\nu - i g \delta_{\nu 0} \bar A_0$. The scalar and
pseudoscalar mesons are encoded in the flavor matrix
$\Sigma = T^a (\sigma_a + i \pi_a)$, where the $T^a$ generate
$U(N_f)$, and $\Sigma_5 = T^a (\sigma_a+i \gamma_5\pi_a)$, see
e.g.\ \cite{Rennecke:2016tkm}.  The couplings of quarks and
mesons to the chemical potential $\mu$ in \Eq{eq:mumat} is achieved by
formally introducing the vector source $C_\nu = \delta_{\nu 0} \mu$
and defining the covariant derivative acting on the meson fields
$\bar D_\nu \Sigma = \partial_\nu \Sigma +[C_\nu,\Sigma]$,
\cite{Kogut:2001id}.

Spontaneous chiral symmetry breaking is captured by the meson
effective potential $\tilde U_k(\Sigma)$,
\begin{align}
\tilde U_k(\Sigma) = U_k(\rho_1,\rho_2) -j_l \sigma_l - j_s\sigma_s - c_A \xi\,.
\end{align} 
It consist of a fully $U(N_f)_L\times U(N_f)_R$ symmetric part $U_k(\rho_1,\rho_2)$, where $\rho_i = \tr\big(\Sigma \Sigma^\dagger\big)^i$ are chiral invariants. The linear terms $j_l \sigma_l$ and $j_s\sigma_s$ explicitly break chiral symmetry and are in one-to-one correspondence to light and strange current quark masses. $\sigma_l \sim \langle \bar l l \rangle$ and $\sigma_s \sim \langle \bar s s \rangle$ are directly related to the light and strange chiral condensates which signal spontaneous chiral symmetry breaking. The last term explicitly breaks $U(1)_A$. This symmetry breaking is a consequence of the axial anomaly and $\xi = \det \Sigma + \det \Sigma^\dagger$ is the instanton-induced 't Hooft determinant \cite{'tHooft:1976up, 'tHooft:1976fv}.

The deconfinement phase transition is captured statistically by
including an effective potential for the gluon background field
$U_\text{glue}(L,\bar L)$, formulated in
terms of the order parameter fields for deconfinement, the Polyakov
loops,
\begin{align}\label{eq:pols}
L = \tr_c \exp(i g \bar A_0 / T)\,,\qquad \bar L = \tr_c [\exp(i g \bar A_0 / T)]^\dagger\,.
\end{align}
They are sensitive to the center symmetry of the $SU(3)$ gauge group and can therefore be used as order parameters for the confinement-deconfinement transition.
In PQM effective models an effective potential for $\bar A_0$ (or equvalently the Polyakov loops) that is fitted to the lattice
equation of state of Yang-Mills theory is used as input and the effects of
dynamical quarks are included through the coupling to the gluonic background
in the quark covariant derivative. For a recent review see \cite{Fukushima:2017csk}. 
We use the parametrization of the Polyakov loop potential 
put forward in \cite{Lo:2013hla}, since it
also captures the lowest-order Polyakov loop susceptibilities which
directly contribute to the particle number susceptibilities
\cite{Fu:2016tey}. The coupling of the gluon background field to the quarks effectively introduces baryons to the system. The reason is that in the presence of $\bar A_0$ the thermal quark number distribution $n_F(E,T,\mu_q)$, where $n_F$ is the ordinary Fermi-Dirac distribution and $\mu_q = \mu_B/3$, is replaced by the modified distribution,
\begin{align}\label{eq:nf}
N_F(E,T, \mu_q; L, \bar L) = \frac{1 + 2 \bar L e^{(E_q-\mu_q)/T}+L e^{2(E_q-\mu_q)/T}}{1 + 3 \bar L e^{(E_q-\mu_q)/T}+3 L e^{2(E_q-\mu_q)/T} + e^{3(E_q-\mu_q)/T}}\,.
\end{align}
In the confined phase with $L, \bar L \approx 0$ it takes the form
$N_F \approx 1/\big\{\!\exp[3 (E_q-\mu_q)/T] +1 \big\}$, which is the
distribution function for a $qqq$-state, a baryon. See
\cite{Fu:2015naa} for a more careful discussion of this behavior. In
the deconfined phase, characterized by $L, \bar L \approx 1$, $N_F$ is identical to the distribution of a
single quark. The terms $\exp[2 (E_q-\mu_q)/T]$ in \Eq{eq:nf} can be
interpreted as intermediate diquark states. So the coupling of the
gluon background field $\bar A_0$ to the quarks leads to a smooth
interpolation between baryons in the hadronic phase and quarks in the
QGP. Even though the effective action in \Eq{eq:ea} only has mesons as
explicit hadronic content, we can still account for baryon
dynamics. Including both a baryon- and a strange chemical potential
allows us to capture the effects of strange and nonstrange baryons
separately.

The intricate interplay between meson, baryon and quark dynamics described in the previous section necessitates a treatment of the system beyond the mean-field approximation. Otherwise, open strange meson fluctuations and in particular kaons for instance, which play a crucial role for strangeness in the hadronic phase, are neglected and the relevant physics can't be captured properly. Furthermore, the couplings of the effective action \eq{eq:ea} are typically strong and we want to compute at finite $\mu_B$. We therefore need a non-perturbative method to compute quantum fluctuations which does not have a sign problem. Here, this is achieved by solving the non-perturbative renormalization group flow equations for the effective action $\Gamma_k$ by means of the functional renormalization group (FRG) \cite{Wetterich:1992yh},
\begin{align}\label{eq:fleq}
\partial_t \Gamma_k = \frac{1}{2} \text{Tr}\, \Big[ \big(\Gamma_k^{(2)}[\Phi] + R_k\big)^{-1}\partial_t R_k\Big]\,,
\end{align}
with $t = \ln (k/\Lambda)$. $\Gamma_k^{(2)}[\Phi]$ is the matrix of
second order functional derivatives of the effective action with
respect to the fields $\Phi = (q,\bar q, \Sigma, L, \bar L)$. $R_k$
implements infrared-regularization at momenta $p^2 \approx k^2$. The trace involves the integration over loop-momenta, the color-,
flavor- and spinor-traces as well as the sum over different particle
species including a minus sign for fermions. Solving \Eq{eq:fleq}
amounts to successively integrating out fluctuations starting from the
initial action $\Gamma_{k=\Lambda}$, with $\Lambda = 900$ MeV in our
case, down to the full quantum effective action $\Gamma_{k=0}$. The
FRG provides a non-perturbative regularization and renormalization
scheme for the resummation of an infinite class of Feynman diagrams,
see, e.g., \cite{Berges:2000ew, Pawlowski:2005xe, Gies:2006wv, Braun:2011pp} for
QCD-related reviews. 

In the following, we use Eqs.\ \eq{eq:ea} and \eq{eq:fleq} to shed more light on the physics discussed in \Sec{sec:strn}. For more details regarding our explicit computations, we refer to \cite{Fu:2018qsk}.

\section{Baryon-strangeness correlations and strangeness neutrality}

\begin{figure}[t]
\centering
\includegraphics[width=.49\textwidth]{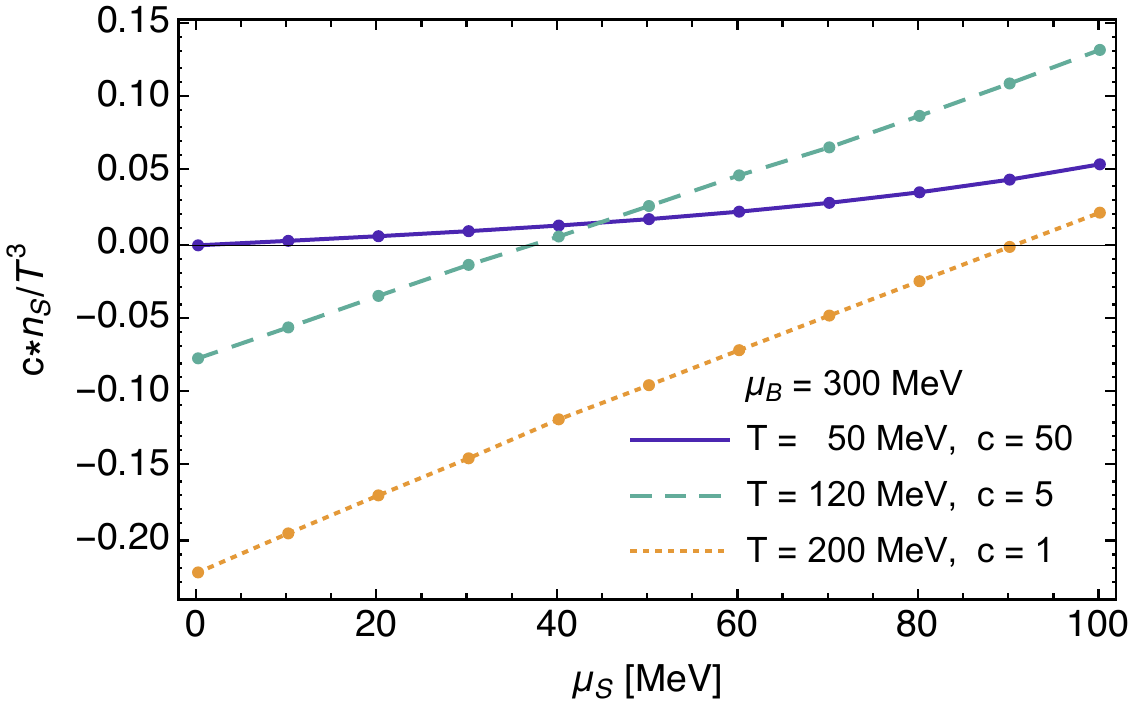}
\hfill
\includegraphics[width=.48\textwidth]{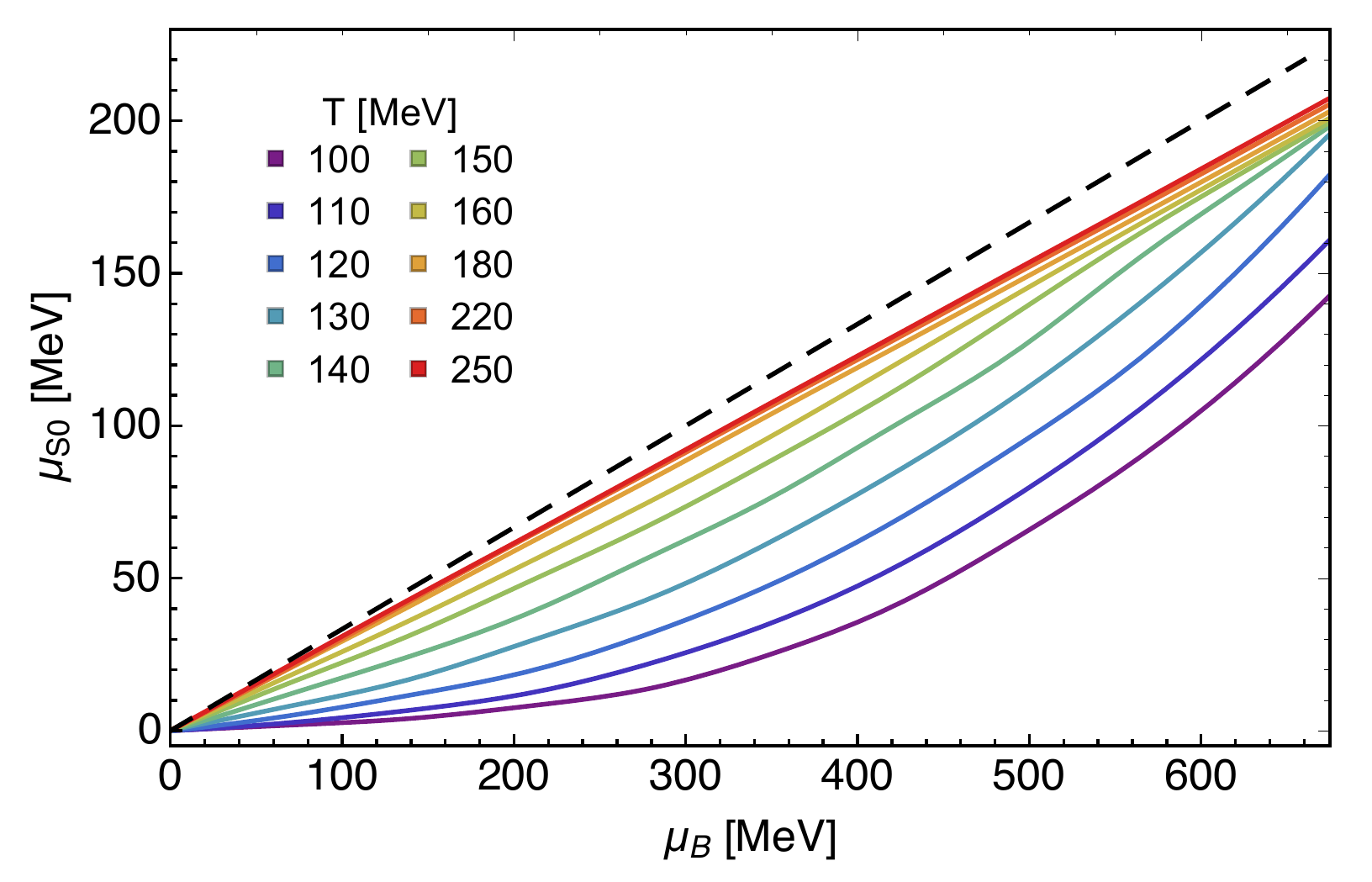}
\caption{\emph{Left}: Strangeness density as a function of $\mu_S$ at $\mu_B = 300$
  MeV for various temperatures. The zero crossings define $\mu_{S0}(T,\mu_B)$. The factor c is just for illustration purposes.
  \emph{Right}: Strangeness chemical potential at strangeness neutrality,
  $\mu_{S0}$, as a function of the baryon chemical potential $\mu_B$
  for various temperatures $T$ (solid lines). T is increasing from
  bottom to top from 100 MeV to 250 MeV. The dashed line corresponds
  to the asymptotic limit of free quarks. In this limit one finds $\mu_{S0} = \mu_B / 3$ according to
  our discussion in \Sec{sec:strn}.}
\label{fig:nsmus0s}
\end{figure}

From the solution of the flow equation \eq{eq:fleq}, we obtain the quantum equations of motion and the thermodynamic potential $\Omega$ as
\begin{align}\label{eq:eos}
\frac{\delta \Gamma_0[\Phi]}{\delta \Phi}\bigg|_{\Phi = \Phi_\text{EoM}} = 0\,, \qquad \Omega = \frac{\Gamma_0[\Phi_\text{EoM}]}{\beta V}\,.
\end{align}
A fist important check of the validity of our model is to compare it to the results of lattice QCD at vanishing and small $\mu_B$. This has been done in \cite{Fu:2018qsk}, where we find very good agreement for various thermodynamic quantities. To extract the strangeness chemical potential at strangeness neutrality, $\mu_{S0}(T,\mu_B)$, we compute the strangeness density $n_S$ for various $T$ and $\mu_B$ as a function of $\mu_S$. $\mu_{S0}$ is then given by the zero-crossings of this function. In the left plot of \Fig{fig:nsmus0s} we show some examples. Note that $n_S$ monotonously increases with $\mu_S$ from negative to positive values at finite $\mu_B$. This is expected since $\mu_B$ increases antistrangeness while $\mu_S$ increases strangeness. We can therefore always find a zero crossing and $\mu_{S0}$ is well-defined.

We use this to extract $\mu_{S0}(T,\mu_B)$. We restrict ourselves to $\mu_B \leq 675$ MeV since our model fails to capture
important qualitative features of the theory at larger $\mu_B$, cf.\ \cite{Fu:2018qsk}. We note that, within the model, the transition is still a crossover in this region. The resulting $\mu_{S0}$ is shown in
the right plot of \Fig{fig:nsmus0s}. The characteristic shape of $\mu_{S0}(\mu_B)$ can
be understood qualitatively from our discussion in \Sec{sec:strn}. \Eq{eq:muscbs} implies that its features are directly linked to $C_{BS}$, which will be discussed in the next paragraph.
We point out that at large temperatures the system undergoes
a crossover to the deconfined phase with a pseudocritical temperature
of $T_d \approx 155$ MeV. Asymptotically,
$\mu_{S0}(\mu_B)$ has to follow \Eq{eq:mus0lim}, indicated by the dashed black line in the figure. However, since the
Polyakov loops are still smaller than unity at $T = 250$ MeV in our
computations, implying that the system is not fully deconfined, this
asymptotic limit is not yet reached here. Still $\mu_{S0}(\mu_B)$ is
approximately linear already at $T \approx 180$ MeV, with a slope only
slightly smaller than $1/3$.

\begin{figure}[t]
\centering
\includegraphics[width=.49\textwidth]{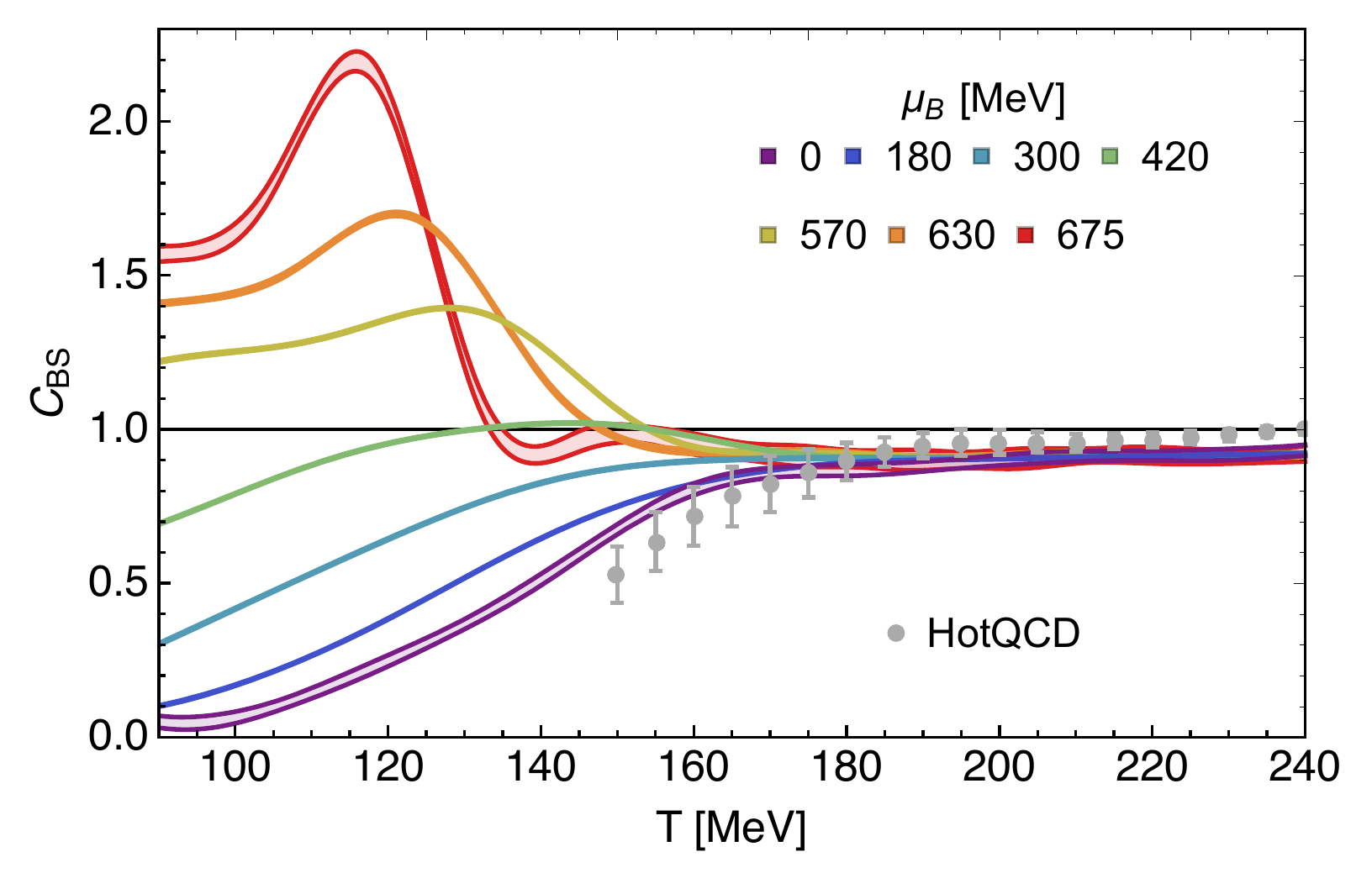}
\hfill
\includegraphics[width=.48\textwidth]{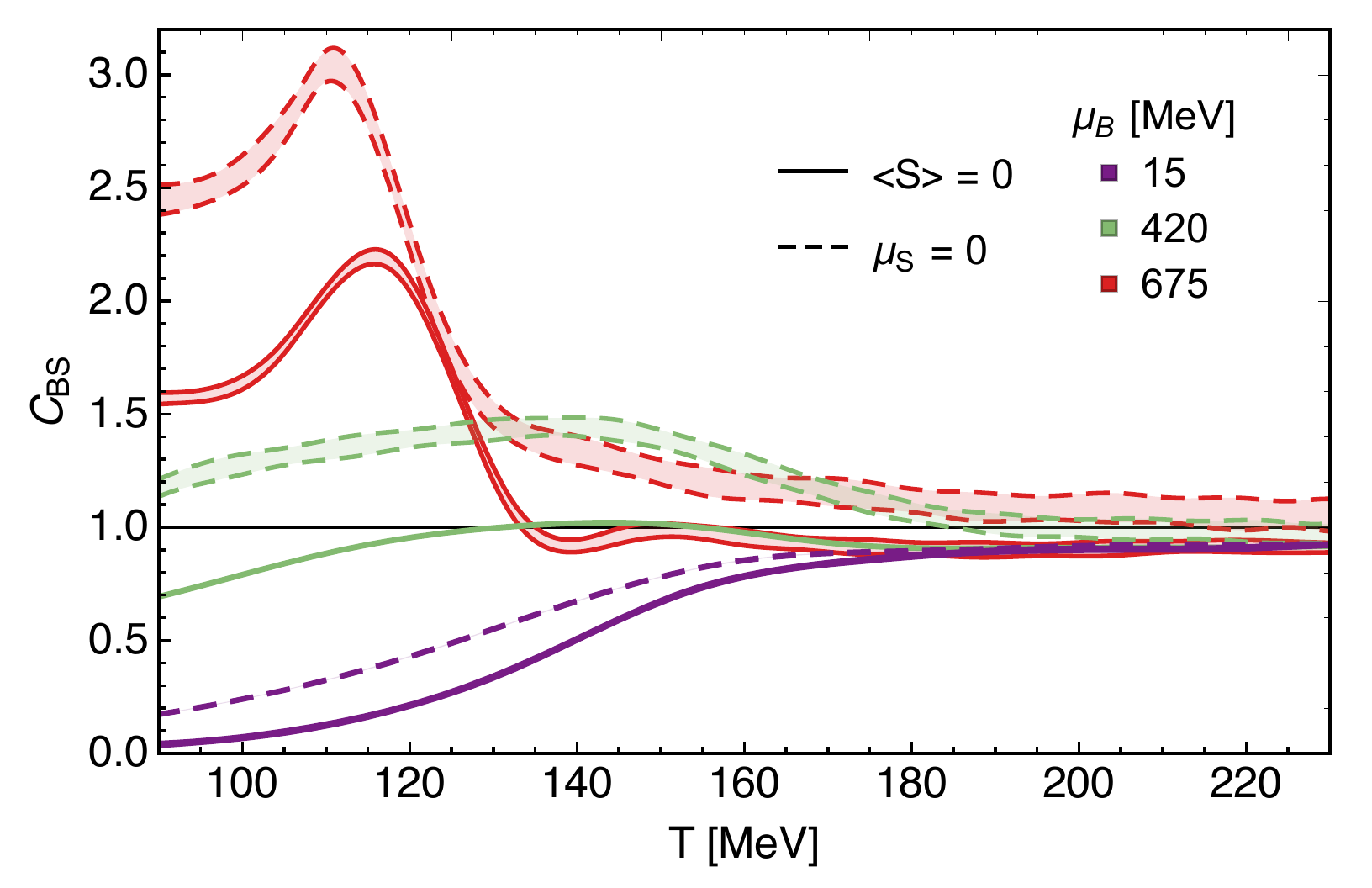}
\caption{\emph{Left}: Baryon-strangeness correlation $C_{BS}$ as a function of
  temperature $T$ for different baryon chemical potential $\mu_B$ at
  strangeness neutrality. $\mu_B$ increases from 0 to 675 MeV from
  bottom to top. At $\mu_B = 0$ we compare to the result of lattice
  QCD \cite{Bazavov:2012jq}. The thin black line indicates the free
  quark limit. The errors reflect the 95\% confidence
  level of a cubic spline interpolation of our numerical data.
  \emph{Right}: Comparison between $C_{BS}$ computed at strangeness
  neutrality, i.e.\ $\mu_S = \mu_{S0}$ (solid lines), and at
  $\mu_S = 0$ (dashed lines), where strangeness conservation is violated at finite $\mu_B$.}
\label{fig:cbsoft}
\end{figure}

Following \Eq{eq:muscbs}, we can extract $C_{BS}$ at strangeness neutrality from the slope of $\mu_{S0}$. This is shown in the left plot of \Fig{fig:cbsoft}. There, $C_{BS}(T,\mu_B,\mu_{S0})$ is shown as a function of $T$ for various $\mu_B$. At $\mu_B = 0$ we compare to lattice results and find good agreement. We have also checked that our results are in good agreement with the hadron resonance gas (HRG) at small temperatures for finite $\mu_B$. This highlights that our model captures the relevant physics quite well. But our results go significantly beyond both the lattice and the HRG. If we focus on the region of small $T$ first, we see that $C_{BS}$ grows with increasing $\mu_B$ and eventually turns from smaller than unity to larger than unity. This is in line with the discussion in \Sec{sec:strn}: At small $\mu_B$ open strange mesons outnumber baryons and thus baryon-strangeness correlations are suppressed. At larger $\mu_B$ baryons, and in particular hyperons, become relevant and start to compensate the (negative) strangeness of kaons, leading to a growing of $C_{BS}$ above unity. This also qualitatively explains the slope of $\mu_{S0}$ turning from smaller to larger than $1/3$ in the right plot of \Fig{fig:nsmus0s}.

With increasing temperature, more baryons can be excited and $C_{BS}$ first grows with $T$. In the deconfined phase it has to approach unity following the discussion of \Eq{eq:cbslim}. Consequently, $C_{BS}$ has to become non-monotonous at large $\mu_B$ where $C_{BS} > 1$. This is clearly seen in \Fig{fig:cbsoft}. In this region, $C_{BS}$ develops a pronounced peak. The peak position coincides with the pseudocritical temperature of the chiral phase transition. Hence, our results demonstrate that $C_{BS}$ not only reflects the intricate interplay of meson, baryon and quark dynamics, but also shows a direct signature of the chiral phase transition. We note that, as opposed to, e.g., the kurtosis of particle number correlations $\sim \chi_4^B/\chi_2^B$, this is not necessarily due to the growing correlation length as the crossover becomes sharper. Since $C_{BS}$ is defined by a ratio of two second-order susceptibilities \eq{eq:cbsdef}, it seems unlikely that it is sensitive to critical physics. But to draw any conclusions, this has to be investigated in more detail. It seems that the origin of the direct sensitivity $C_{BS}$ to the QCD phase transition is due to the distinct difference between the dynamics of the dominant degrees of freedom in the hadronic and the QGP phase, and not necessarily due to critical physics.

To asses the relevance of strangeness neutrality on baryon-strangeness correlations, we compare our results for $C_{BS}$ at strangeness neutrality to those obtained at $\mu_S = 0$ in the right plot of \Fig{fig:cbsoft}. Since there is no relation between $\mu_S$ and $C_{BS}$ for $\mu_S = 0$, we have to compute it from \Eq{eq:cbsdef} directly in this case. We see that $C_{BS}$ is significantly larger at $\mu_S = 0$ compared to at strangeness neutrality. This is due to the fact that increasing $\mu_{S}$ compensates for the strangeness generated by increasing $\mu_B$. If $\mu_S$ is not adjusted to enforce strangeness neutrality, more strange baryons can be excited. We see that taking into account the conservation of net-strangeness is crucial for an accurate description of the baryon-strangeness correlations present in heavy-ion collisions.

\begin{figure}[t]
\centering
\includegraphics[width=.5\textwidth]{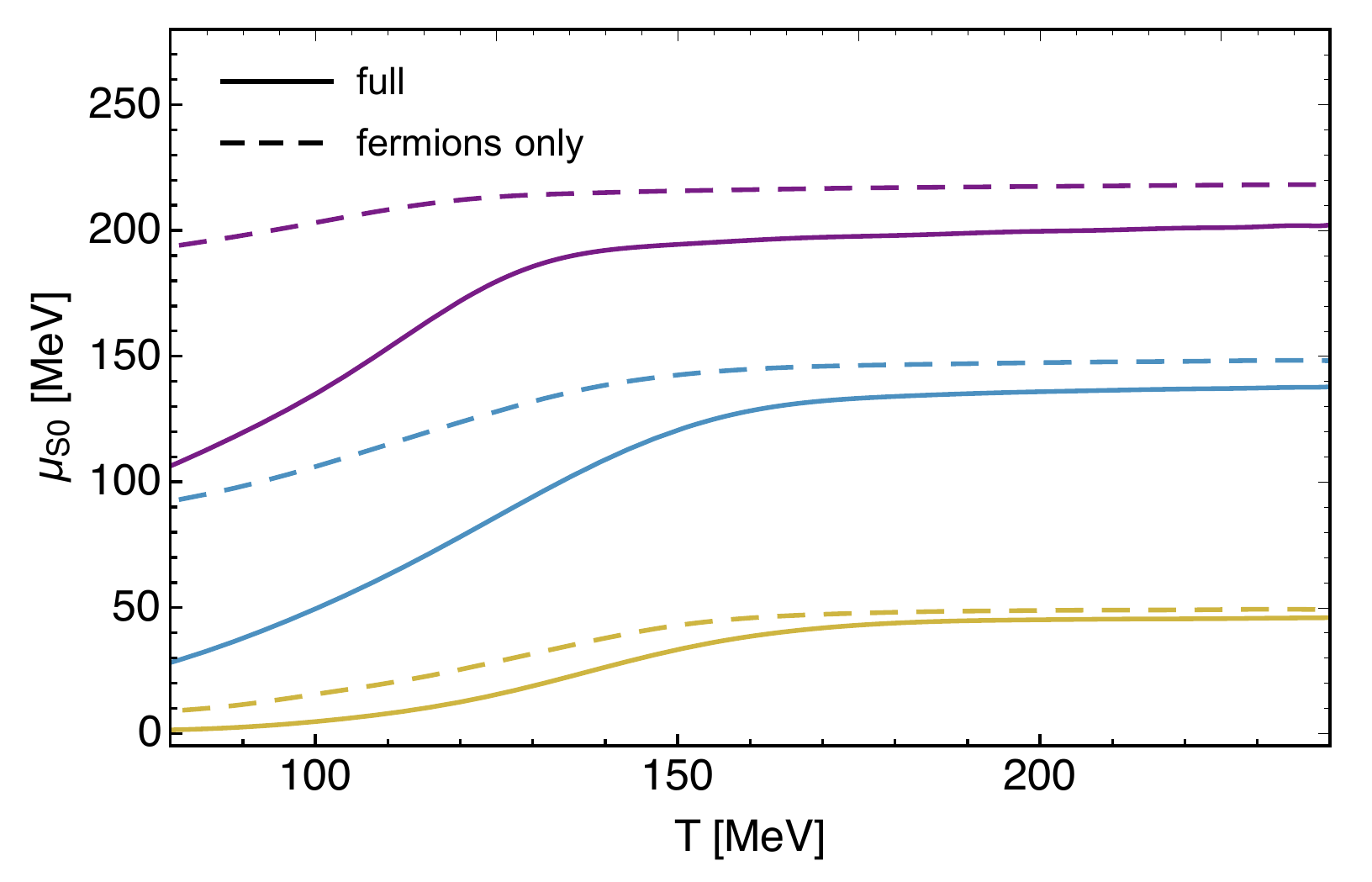}
\caption{Comparison between our full result for $\mu_{S0}$ (solid
  lines) and \Eq{eq:fuku} (dashed lines) for $\mu_B = 150,\, 450$ and
  $660$ MeV (from bottom to top).}
\label{fig:fukurel}
\end{figure}

We have argued that it is crucial to treat the system beyond mean-field approximations in order to capture the important dynamics of open strange mesons. To corroborate this  statement, we compare our result for $\mu_{S0}$ with the result obtained within a mean-field approximation. To this end, we use the relation:
\begin{align}\label{eq:fuku}
\mu_{S0}\Big|_\text{fermions}(T,\mu_B) \approx \frac{\mu_B}{3} - \frac{T}{2} \ln\left[ \frac{\bar L(T,\mu_B)}{L(T,\mu_B)} \right]\,,
\end{align}
which can be derived from the fermion contribution to the flow equation \label{eq:fleq}, assuming that the (ratio of) the Polyakov loops is independent of $\mu_S$. This equation has been derived in \cite{Fukushima:2009dx} and it has been shown to be about 3\% accurate for a mean-field study. A comparison to our FRG result is a good indication for the relevance of meson fluctuations. This is shown in \Fig{fig:fukurel}, where we show $\mu_{S0}$ as a function of $T$ for different $\mu_B$. We see that fluctuation effects beyond mean-field are crucial. The competing effects on strangeness between open strange mesons and strange baryons is not captured due to the missing meson fluctuations in the mean-field approximation.

\section{QCD phase structure and thermodynamics}

\begin{figure}[t]
\centering
\includegraphics[width=.32\textwidth]{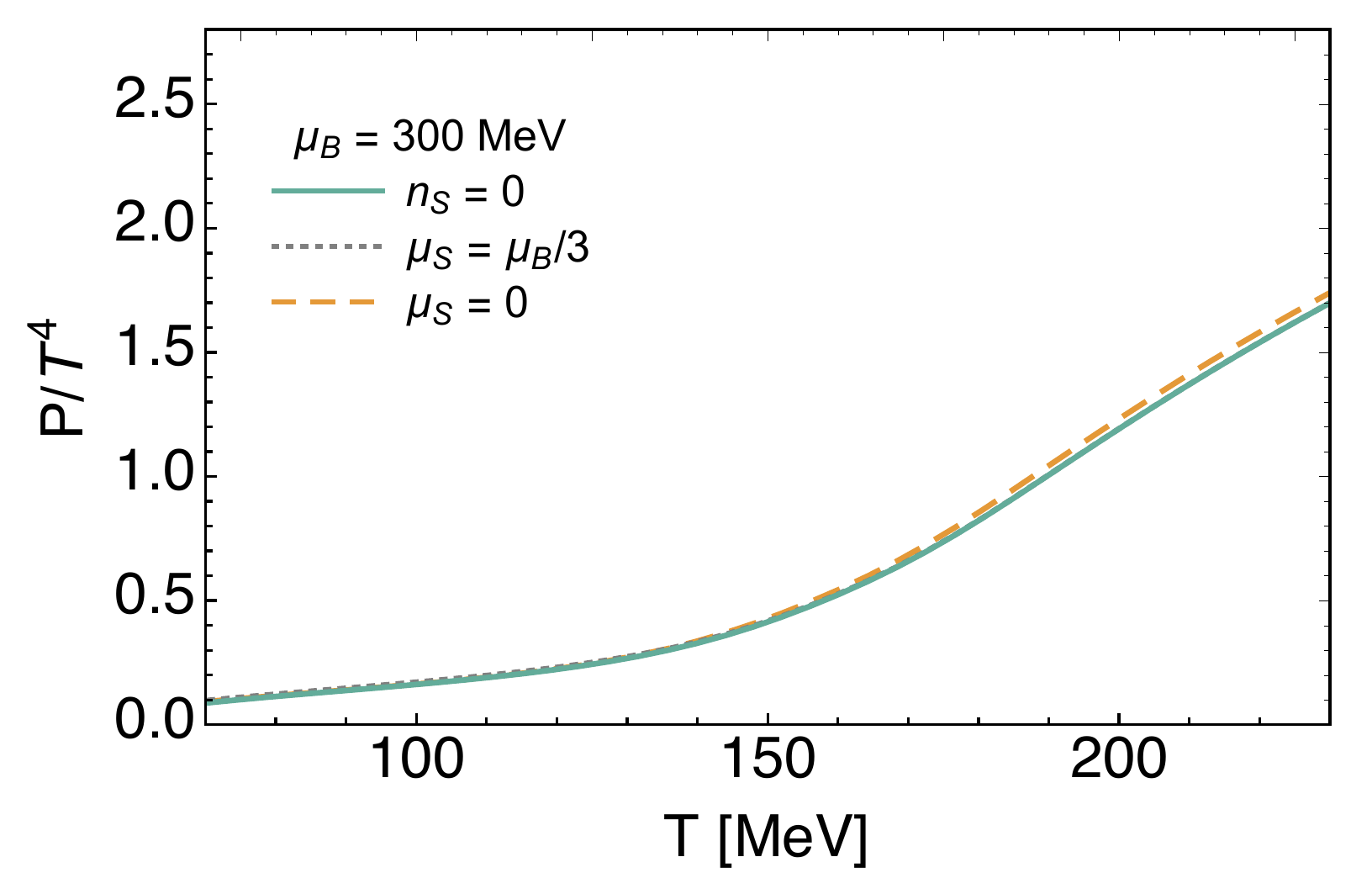}
\includegraphics[width=.32\textwidth]{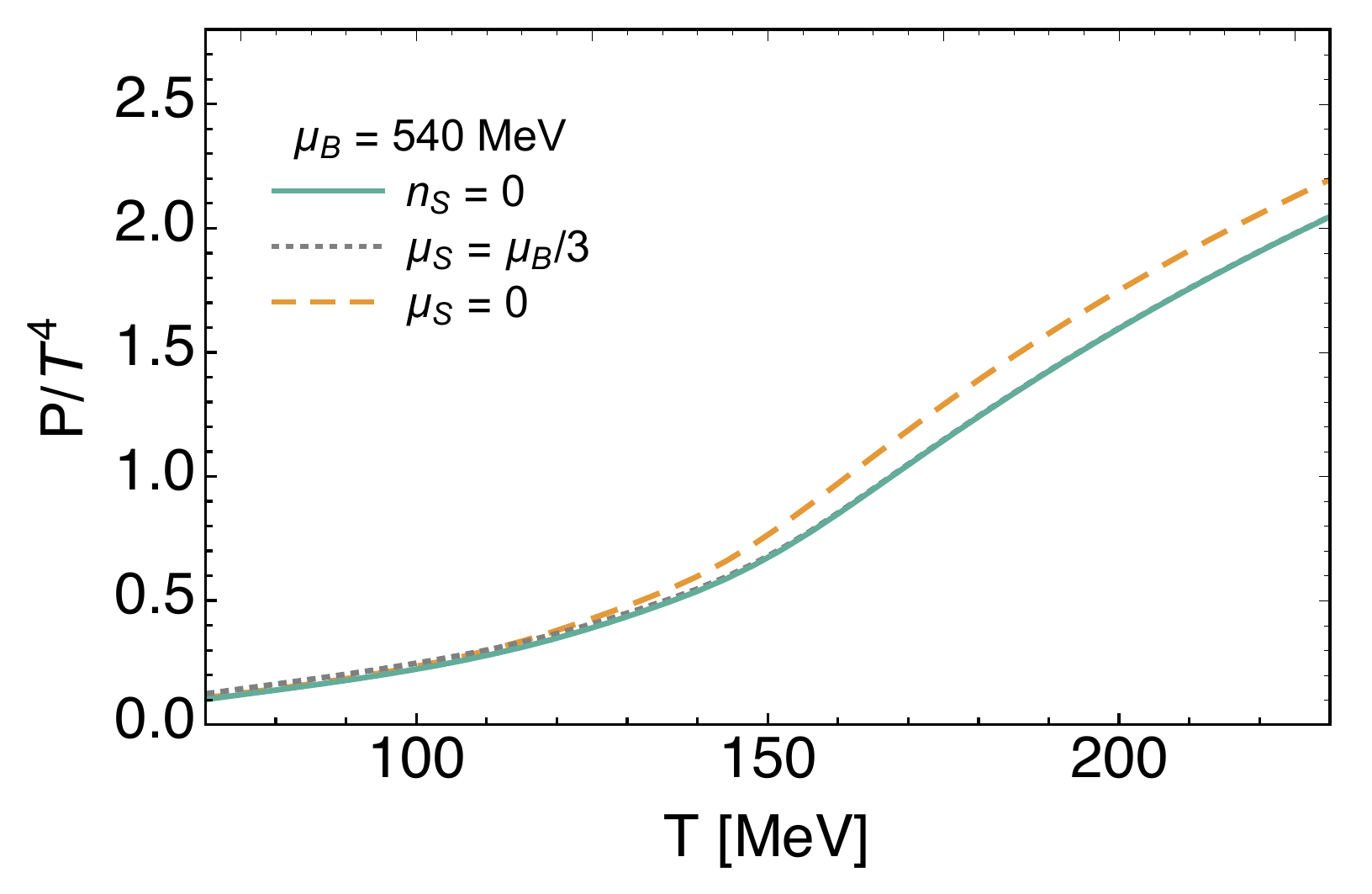}
\includegraphics[width=.32\textwidth]{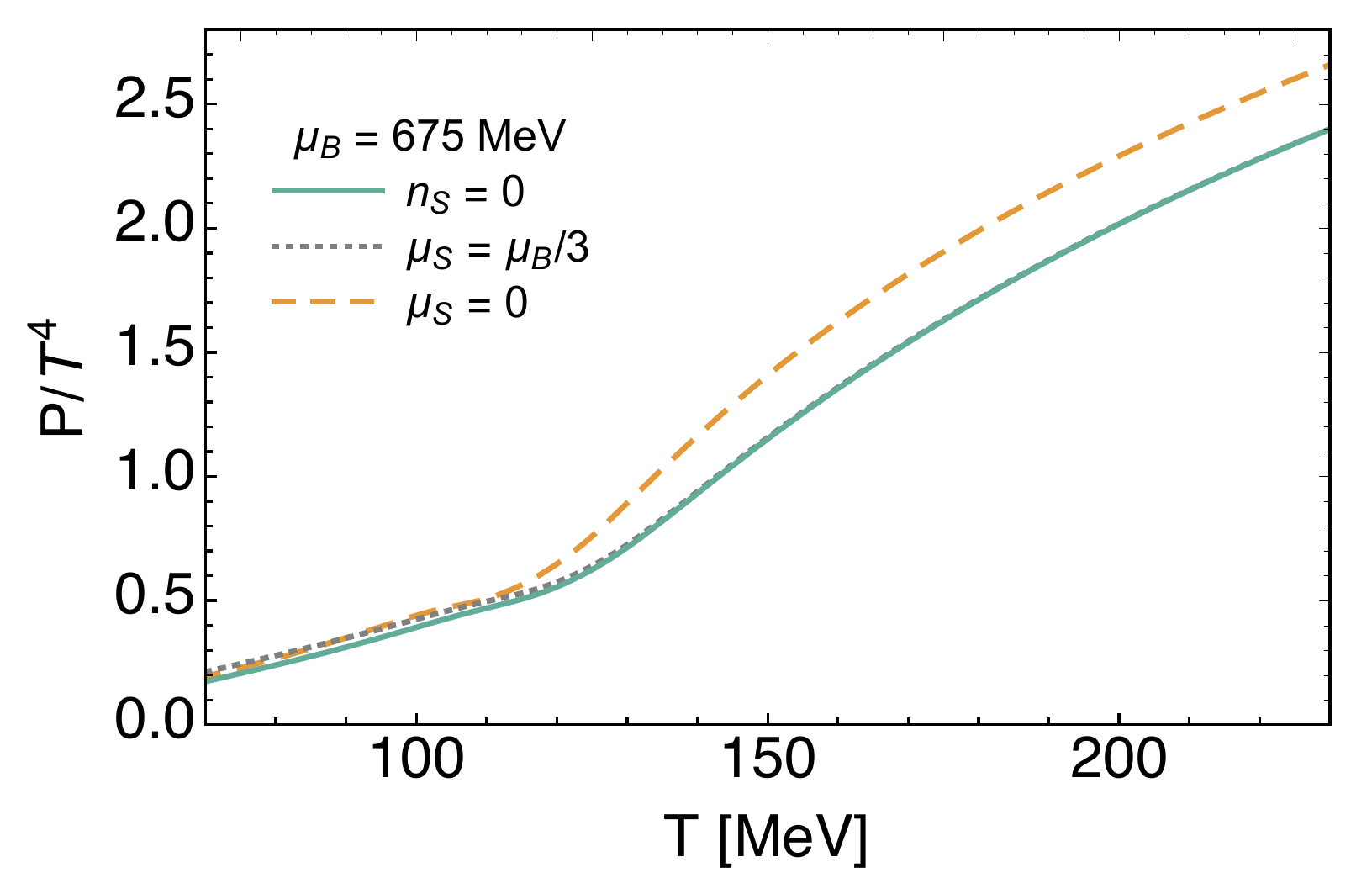}
\includegraphics[width=.32\textwidth]{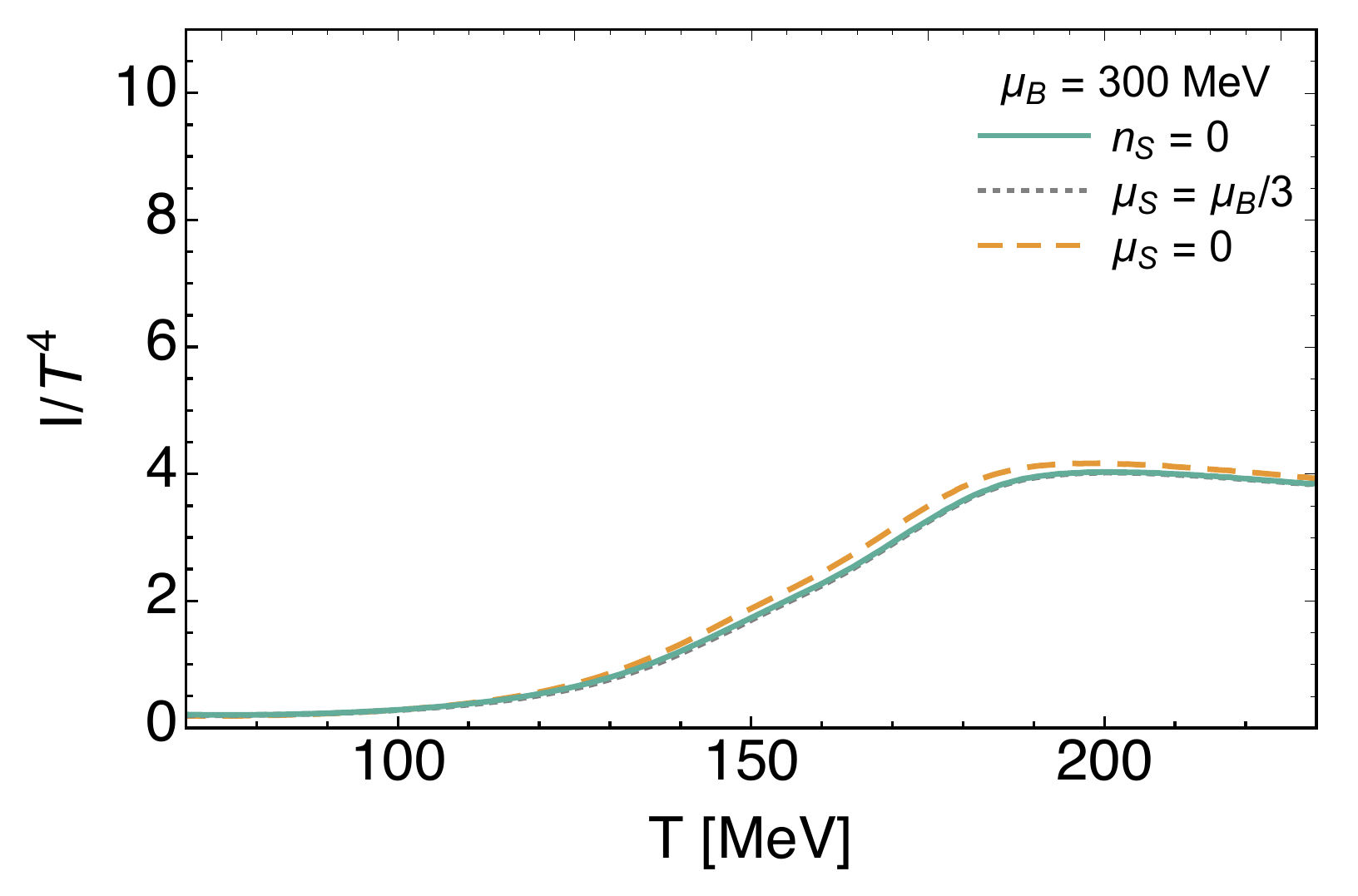}
\includegraphics[width=.32\textwidth]{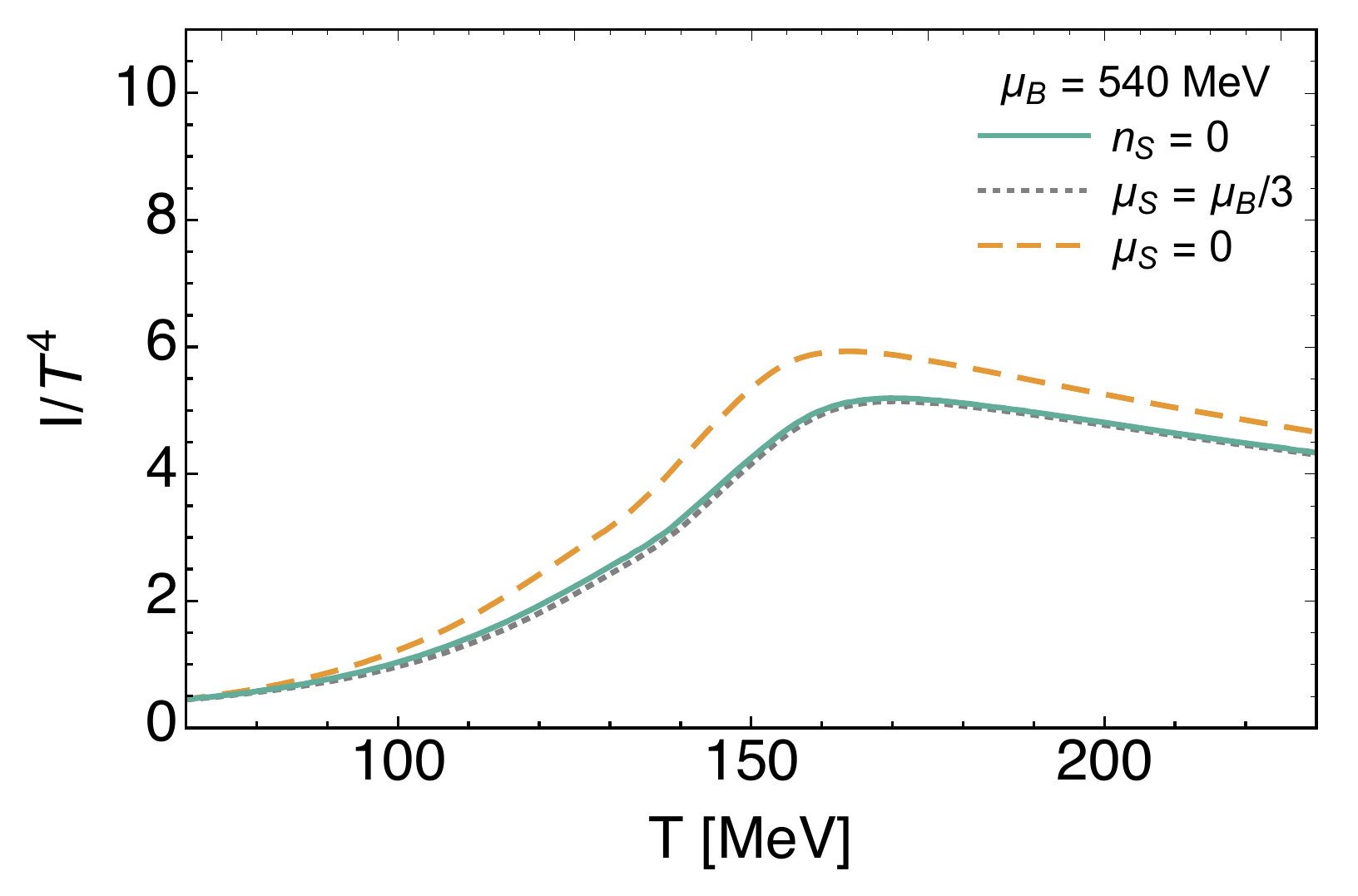}
\includegraphics[width=.32\textwidth]{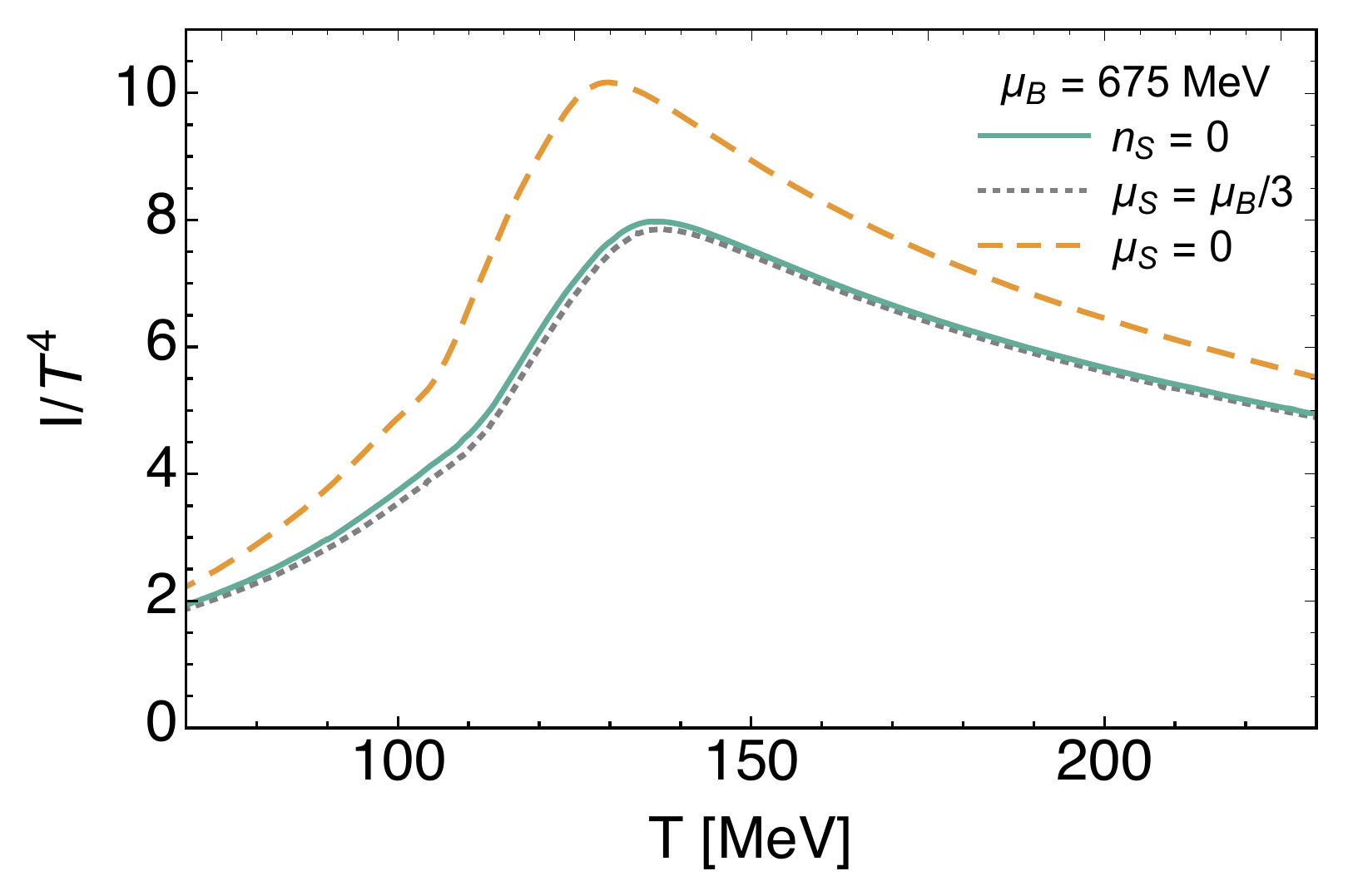}
\caption{Comparison between the pressure (first row) and the trace anomaly (second row) at strangeness neutrality (solid blue line), at $\mu_S = 0$ (dashed orange line) and at $\mu_S = \mu_B/3$ (dotted gray line) for various $\mu_B$ as functions of $T$.}
\label{fig:tdcomp}
\end{figure}

Here, we investigate the impact of net-strangeness conservation on the phase structure and thermodynamics. From the thermodynamic potential $\Omega$ in \Eq{eq:eos}, we can extract all thermodynamic quantities, e.g.\ the pressure $p$, the entropy $s$, the energy density $\epsilon$, the trace anomaly $I$,
\begin{align}\label{eq:thermo}
\begin{split}
p = - \Omega\,,\qquad s = \frac{\partial p}{\partial T}\,, \qquad \epsilon = -p + T s + \mu_B n_B + \mu_S n_S\,, \qquad I = \epsilon - 3p\,,
\end{split}
\end{align}
and, of course, the generalized susceptibilities in \Eq{eq:sus}. We compare these quantities for three different cases: At strangeness neutrality $n_S = 0$ with $\mu_S = \mu_{S0}(T,\mu_B)$, at vanishing strangeness chemical potential $\mu_S = 0$, and at $\mu_S = \mu_B/3$. The latter choice is motivated by having a vanishing strange quark chemical potential, cf.\ \eq{eq:mumat}. Deep in the deconfined phase this is identical to $\mu_{S0}$ according to \Eq{eq:mus0lim}. We show our results for the pressure and the trace anomaly in \Fig{fig:tdcomp}. At small $\mu_B$ there is no significant difference between all these choices since the equation of state is not very sensitive to chemical potential effects in this regime. The difference between $\mu_S = \mu_{S0}$ and $\mu_S = 0$ becomes significant already at moderate chemical potentials. For the pressure, the reason is that the larger $\mu_B$, the more steeply it rises with $T$ above the phase transition since the phase transition becomes sharper. A finite $\mu_{S0}$ removes part of the baryon dynamics from the system and thus delays this effect. As expected, the choice $\mu_S = \mu_B/3$ is in perfect agreement with the strangeness neutral result at large temperatures. Even at small temperatures, we only see small deviations. In this region, the pressure is not very sensitive to any variations in $\mu_S$. In the hadronic phase pion fluctuations dominate the pressure. The trace anomaly is more sensitive to strangeness effects. The reason is that the baryon and strangeness densities directly contribute to $I$. With increasing $\mu_S$, also the baryon density is suppressed. At strangeness neutrality with $n_S = 0$ this leads to a significant suppression of the trace anomaly at large $\mu_B$ as compared to $\mu_S = 0$. Note that in both cases, the $\mu_S n_S$ contribution to $I$ in \Eq{eq:thermo} vanishes. Interestingly, also here $\mu_S = \mu_B/3$ is a good approximation to the result at strangeness neutrality over the full range of temperatures. We have shown in \cite{Fu:2018qsk} that this is not the case for the speed of sound, which we do not discuss here.

\begin{figure}[t]
\center
\includegraphics[width=.483 \textwidth]{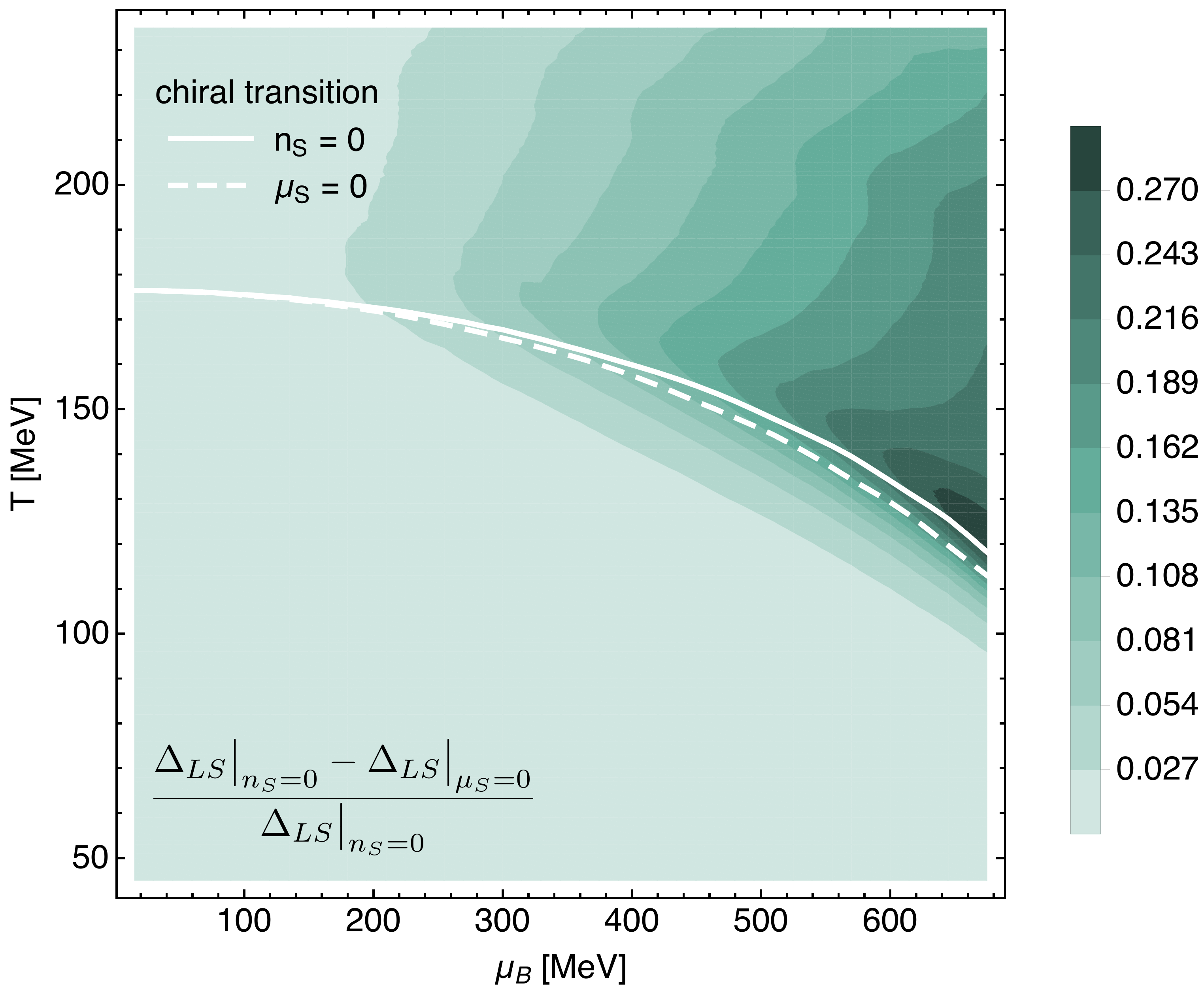}
\hfill
\includegraphics[width=.49 \textwidth]{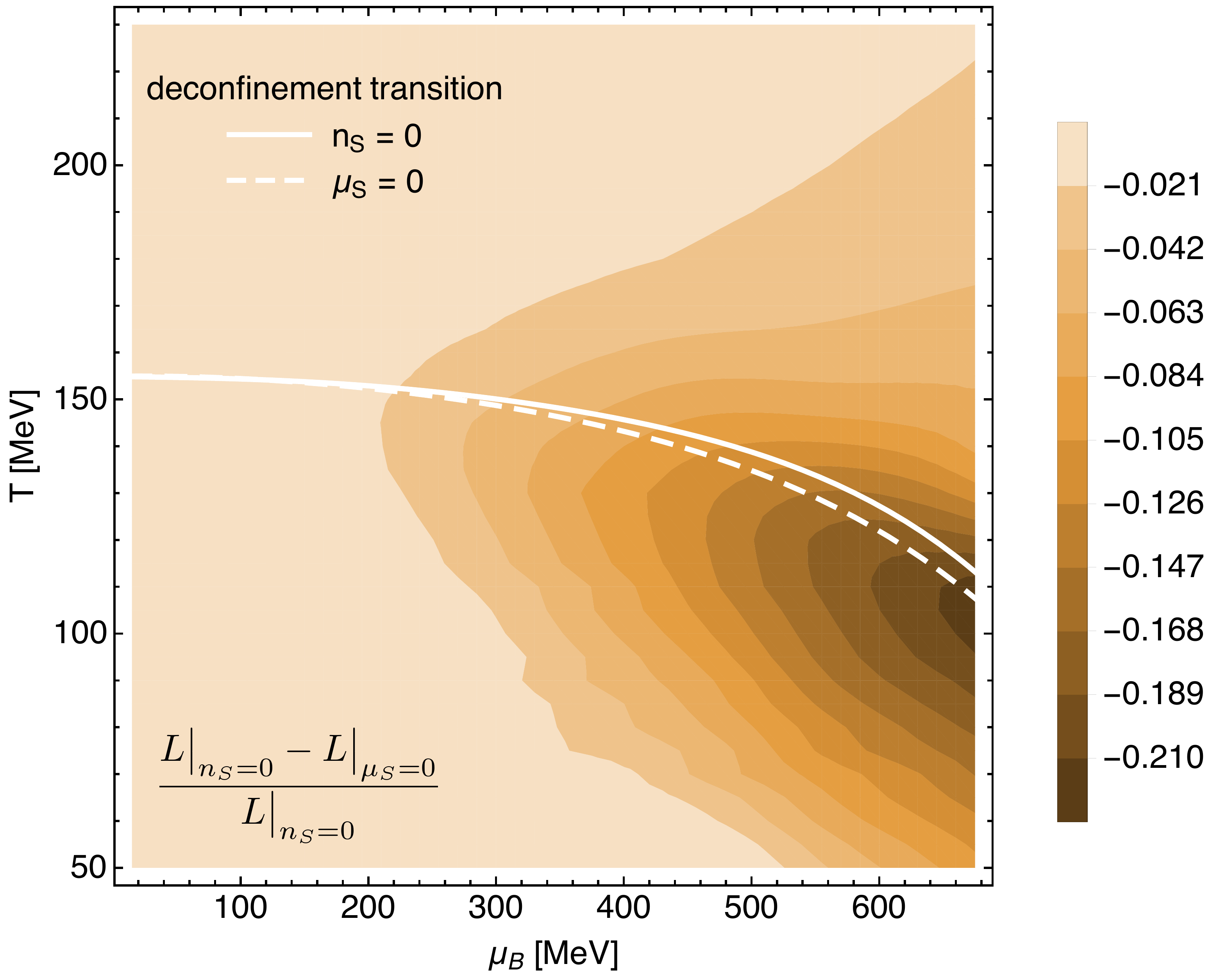}
\caption{Chiral and deconfinement phase diagrams projected onto the $(T,\,\mu_B)$-plane. \emph{Left}: Relative error of the subtracted condensate for strangeness neutrality and $\mu_S = 0$. The solid  and dashed lines indicate the chiral phase boundary as defined by the inflection point of $\Delta_{LS}(T)$ for $n_S=0$ and $\mu_S=0$ respectively. \emph{Right}: The same for the Polyakov loop. Here, the solid and dashed lines indicate the deconfinement phase boundary as defined by the inflection point of $L(T)$ for $n_S=0$ and $\mu_S=0$.}
\label{fig:pb}
\end{figure}

To study the effects on the phase structure itself, we look at the chiral and deconfinement phase transitions separately. In the crossover region, the result for the phase transition temperature is not unique. It depends of the definition of the pseudocritical temperature, see e.g.\ \cite{Pawlowski:2014zaa}. For the order parameter of the chiral phase transition we use the subtracted condensate,
\begin{align}\label{eq:subcond}
\Delta_{LS} = \frac{\big(\sigma_l - \frac{j_l}{j_s} \sigma_s\big)\big|_{T}}{\big(\sigma_l - \frac{j_l}{j_s} \sigma_s\big)\big|_{T=0}}\,.
\end{align}
For the confinement-deconfinement phase transition we use the Polyakov loop $L$ in \Eq{eq:pols}. Note that the antiloop $\bar L$ yields exactly the same results for the phase transition as $L$. In any case, we define the pseudocritical temperature as the  inflection point of the order parameter as a function of $T$ for any $\mu_B$. To study the effect of strangeness conservation, we again compare the case $\mu_S = \mu_{S0}$ and $\mu_S = 0$. Since both transitions are crossovers, a meaningful assessment requires the comparison of the global shape of the order parameters. To this end we also compute the relative differences of the order parameters $\Delta_{LS}(T,\mu_B)$ and $L(T,\mu_B)$,
\begin{align}\label{eq:odiff}
\frac{\Delta_{LS}\big|_{n_S=0}-\Delta_{LS}\big|_{\mu_S=0}}{\Delta_{LS}\big|_{n_S=0}}\,, 
\qquad
 \frac{L\big|_{n_S=0}-L\big|_{\mu_S=0}}{L\big|_{n_S=0}}\,.
\end{align}
The phase diagrams for the chiral and deconfinement transitions in the $(T,\, \mu_B)$-plane are shown in \Fig{fig:pb}. Note that these diagrams are projections onto this plane. The results at strangeness neutrality actually extend additionally along the $\mu_S$ axis. The solid and dashed white lines show the pseudocritical temperatures for $\mu_S = \mu_{S0}$ and $\mu_S = 0$ respectively. The density profile shows the relative difference of the order parameters in \Eq{eq:odiff}. The pseudocritical lines at strangeness neutrality are always above the ones at $\mu_S = 0$ and the difference grows with increasing $\mu_B$. This delay of the phase transition to lager $\mu_B$ at a given $T$ has the same origin as our findings for the pressure. Baryon effects (relative to meson effects), which are responsible for the sharpness of the crossover and also the bending of the phase boundary towards the $\mu_B$-axis with increasing $\mu_B$, are partly suppressed at finite $\mu_S$ since it effectively removes strange baryon contributions. However, even at the larges baryon chemical potential considered here, $\mu_B = 675\,\text{MeV}$, this only has roughly a 5\% effect on the pseudocritical temperatures.

This is drastically different if one looks at the overall difference of the order parameters. There, the effect on the chiral transition is up to about 30\% and on the deconfinement transition up to about 20\%. The relation $\Delta_{LS}\big|_{n_S=0} \geq \Delta_{LS}\big|_{\mu_S=0}$ holds for
all $T$ and $\mu_B$ considered here, so the chiral condensate starts
melting at larger $T$ and more slowly at strangeness neutrality. The effect is largest above the pseudocritical line. In the hadronic phase, the differences are very small. For the Polyakov loops we always find $L\big|_{n_S=0} \leq L\big|_{\mu_S=0}$. Here, the differences are larges below the pseudocritical line in the hadronic phase. Recalling that the QGP phase is characterized by \emph{restored} chiral symmetry and \emph{broken} center symmetry, our results show that the phase phase diagram in the region of restored symmetry is most sensitive to strangeness conservation.

\begin{figure}[t]
\centering
\includegraphics[width=.6\textwidth]{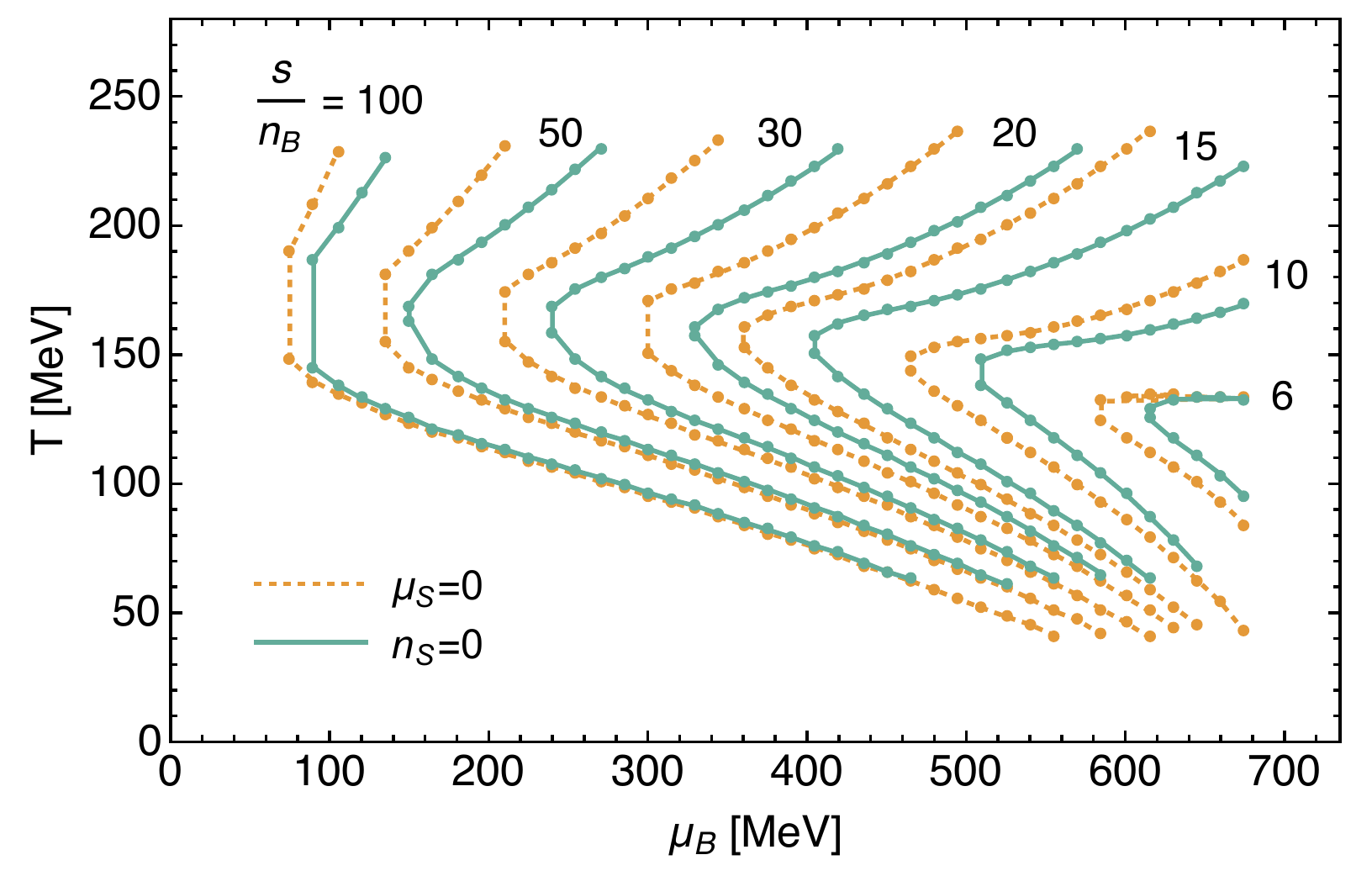}
\caption{Isentropes in the phase diagram projected onto the $(T,\,\mu_B)$-plane.}
\label{fig:isen}
\end{figure}

While we didn't push our computation to the CEP of the model, we can still speculate on the effect of strangeness neutrality on its location. We note that the crossover is already very sharp at $\mu_B = 675\,\text{MeV}$, so it is unlikely that the CEP is at much larger $\mu_B$. Since the phase transition is of second order the CEP, the critical temperature is uniquely defined. The phase transition can then be read-off from the global shape of the order parameters. We have shown that at large $\mu_B$ strangeness has a large effect on the order parameters. Taking strangeness neutrality into account could therefore potentially also have a $\sim30\%$ effect on the location of the CEP. Following our discussion above, we expect it to move to larger $\mu_B$ as compared to $\mu_S = 0$. We emphasize that we do not expect the present model to give an accurate prediction of the location of the CEP. However, the physical effects observed here will certainly also be relevant in a more quantitative study of the phase diagram.

Finally, we study isentropes in the phase diagram. Their relevance derives from the fact that the QGP evolves hydrodynamically at the later stages of the heavy-ion collision. It has been shown that this QGP fluid is almost ideal with a very small shear viscosity over entropy density ratio, see e.g.\ \cite{Gale:2013da}. This implies that the QGP evolves through the phase diagram close to the isentropes, defined by a fixed ratio of entropy and baryon number, $s/n_B$. Again, we compare $\mu_S = \mu_{S0}$ and $\mu_S = 0$. The results are shown in \Fig{fig:isen}. The isentropes show a very characteristic behavior: They bend in the phase transition region and the sharper the transition, the sharper the bending. 
Interestingly, in studies of the isentropes within two-flavor PQM models, this only occurs for small $s/n_B$ \cite{Skokov:2010uh}. Hence, the sensitivity of the isentropes to the phase transition at large $s/n_B$ can be attributed to strangeness.

The behavior of the isentropes in
the hadronic phase is dictated by the Silver-Blaze property of QCD \cite{Cohen:2003kd}. At
$T = 0$ and $\mu_B \lesssim m_N$, where $m_N$ is the nucleon mass, the baryon number has to
vanish. Hence, the isentropic curves bend towards larger $\mu_B$ with
decreasing $T$. The difference between $n_S = 0$ and $\mu_S=0$ is
small at small temperatures because the lightest baryonic resonance
does not carry strangeness. Furthermore, the entropy density grows with $T$. The baryon
number, on the other hand, has a maximum at the chiral phase
transition and slowly decreases with increasing temperature above
$T_\chi$. Hence, the isentropes bend towards larger $\mu_B$ with
increasing $T$ above the phase transition. The regions where the
isentropes turn therefore clearly indicate the transition
region. Since the baryon number at strangeness neutrality is
systematically smaller than for $\mu_S = 0$ at a given $\mu_B$, the
bending of the isentropes above the phase transition is stronger at
strangeness neutrality. We also find that the isentropes at
strangeness neutrality are systematically shifted towards larger $\mu_B$. This can also be understood from the fact that the
baryon number decreases with increasing $\mu_S$. This effect dominates
over the corresponding effect on the entropy density (which behaves
very similar to the pressure in \Fig{fig:tdcomp}). The overall effect of strangeness conservation on the evolution of the QGP through the phase diagram is significant.

\section{Conclusions}

We have demonstrated that particle number conservation relevant for heavy-ion collisions has significant effects on the phase structure and thermodynamics of QCD. On the example of net-strangeness conservation, we have shown that there is an intricate interplay between hadronic and partonic sources of strangeness which is highly sensitive to the composition of QCD matter. One consequence of this is that baryon-strangeness correlations carry a direct signature of the QCD phase transition. Our results show that it is indispensable to take strangeness neutrality into account for an accurate description of strongly interacting matter in heavy-ion collisions.

\acknowledgments{
F.R.\ thanks the organizers and participants of the \emph{CPOD 2018} for an inspiring conference. His work is supported by the Deutsche Forschungsgemeinschaft (DFG) through grant \mbox{RE 4174/1-1} and the Nuclear Theory Group of the Brookhaven National Laboratory. W.F. is supported by the National Natural Science Foundation of China under Contracts No.\ 11775041. J.M.P.\ acknowledges support by the ExtreMe Matter Institute
(EMMI), the grant BMBF 05P12VHCTG and the DFG Collaborative Research Centre "SFB 1225 (ISOQUANT)".
}

\vspace{-0.2cm}
\bibliographystyle{JHEP}
\bibliography{qcd-phase}

\end{document}